\documentstyle [12pt] {article}

\catcode`@=12
\begin{document}
\thispagestyle{empty}
\begin{flushright} UCRHEP-T136\\November 1994\\
\end{flushright}
\vspace{0.5in}
\begin{center}
{\Large \bf Simple Radiative Neutrino Mass Matrix\\
for Solar and Atmospheric Oscillations\\}
\vspace{1.5in}
{\bf Ernest Ma\\}
\vspace{0.3in}
{\sl Department of Physics\\}
{\sl University of California\\}
{\sl Riverside, California 92521\\}
\vspace{1.5in}
\end{center}
\begin{abstract}\
A simple $3 \times 3$ neutrino Majorana mass matrix is proposed to
accommodate both the solar and atmospheric neutrino deficits.  This
scenario can be realized naturally by a radiative mechanism
for the generation of neutrino masses.  It also goes together naturally
with electroweak baryogenesis and cold dark matter in a specific model.
\end{abstract}

\newpage
\baselineskip 24pt

There is now a good deal of evidence from different experiments that
there exists a solar neutrino deficit\cite{1,2,3,4} as well as mounting
evidence for an atmospheric neutrino deficit.\cite{5,6}  In terms of
neutrino oscillations, the former (latter) is an indication that
$\nu_e$ ($\nu_\mu$) is not a mass eigenstate.\cite{7,8}  A popular
approach to the neutrino-mass problem is the seesaw mechanism,\cite{9}
in which case $m_{\nu_l}$ is naively expected to be proportional to
$m_l^2$, where $l = e, \mu, \tau$, and the mixing angles are assumed to
be small, in analogy with what is observed in the quark sector.  However,
that is not the only, nor necessarily the most natural, possibility.
In this paper, a very different form of the neutrino mass matrix will
be proposed.  It is simple and can be realized naturally by a
radiative mechanism for the generation of Majorana neutrino masses.
It also fits very well into the framework of a recently proposed doublet
Majoron model\cite{10,11} which allows for the generation of baryon number
during the electroweak phase transition as well as having $\nu_\tau$ as
the late decaying particle for a consistent interpretation that
the missing mass of the Universe is all cold dark matter.

Consider the following $3 \times 3$ Majorana mass matrix for the states
$\nu_e, \nu_\mu$, and $\nu_\tau$ (or $\nu_S$, a hypothetical singlet neutrino):
\begin{equation}
{\cal M}_\nu = \left[ \begin{array}{c@{\quad}c@{\quad}c} \epsilon_1 &
\epsilon_4 & m \cos \theta \\
\epsilon_4 & \epsilon_2 & m \sin \theta \\ m \cos \theta &
m \sin \theta & \epsilon_3 \end{array} \right],
\end{equation}
where
\begin{equation}
\epsilon_{1,2,3,4} << m.
\end{equation}
Let the mass eigenstates be denoted by $n_{1,2,3}$, then the corresponding
mass eigenvalues are
\begin{eqnarray}
m_1 &\simeq& m + {1 \over 2} (\epsilon_1 \cos^2 \theta + \epsilon_2
\sin^2 \theta + \epsilon_3 + \epsilon_4 \sin 2 \theta), \\ m_2 &\simeq& - m +
{1 \over 2} (\epsilon_1 \cos^2 \theta + \epsilon_2 \sin^2 \theta +
\epsilon_3 + \epsilon_4 \sin 2 \theta), \\ m_3 &\simeq& \epsilon_1 \sin^2
\theta + \epsilon_2 \cos^2 \theta - \epsilon_4 \sin 2 \theta,
\end{eqnarray}
and
\begin{eqnarray}
\nu_e &\simeq& {1 \over \sqrt 2} (n_1 - n_2) \cos \theta - n_3 \sin \theta,
\\ \nu_\mu &\simeq& n_3 \cos \theta + {1 \over \sqrt 2} (n_1 - n_2)
\sin \theta, \\ \nu_\tau (\nu_S) &\simeq& {1 \over \sqrt 2} (n_1 + n_2).
\end{eqnarray}
{}From Eqs. (3) - (5), we see that
\begin{eqnarray}
\Delta m_{12}^2 &\simeq& 2 m (\epsilon_1 \cos^2 \theta + \epsilon_2
\sin^2 \theta + \epsilon_3 + \epsilon_4 \sin 2 \theta) \nonumber \\ &<<& m^2
\simeq \Delta m_{13}^2 \simeq \Delta m_{23}^2.
\end{eqnarray}
This means that $\nu_\mu - \nu_e$ oscillations are governed by $m^2$
and $\sin^2 2 \theta$, which can be chosen to be about $10^{-2} {\rm eV}^2$
and 0.5 respectively\cite{12} to account for the atmospheric neutrino
data.\cite{5,6}  As for the solar neutrino deficit, the
$\nu_e$ flux is first diminished by its rapid oscillation into $\nu_\mu$
to $(1- {1 \over 2} \sin^2 2 \theta)$ of its initial value, then the
oscillation into $\nu_\tau$ (or $\nu_S$) with $\sin^2 2 \theta_{12} = 1$
and $\Delta m_{12}^2$ of about $10^{-10} {\rm eV}^2$ for the vacuum
oscillation solution\cite{13} reduces it further\cite{14} to what is
observed.\cite{1,2,3,4}
Matter-enhanced oscillations\cite{15} are not possible here because
the mixing is maximum, {\it i.e.} $\theta_{12} = \pi/4$.

The above discussion shows that as long as the $\nu_e - \nu_\tau$ (or
$\nu_e - \nu_S$) and $\nu_\mu - \nu_\tau$ (or $\nu_\mu - \nu_S$)
entries of the $3 \times 3$ Majorana neutrino mass
matrix are much greater than all other entries, the resulting mass
eigenstates will be such that a linear combination of $\nu_e$ and
$\nu_\mu$ pairs up with $\nu_\tau$ (or $\nu_S$) to form a pseudo-Dirac
neutrino, {\it i.e.} an equal (or almost equal) admixture of two nearly
degenerate Majorana neutrinos.  With suitable values for the two large
entries and a general magnitude for the small ones, both the solar and
atmospheric neutrino deficits are explained.  The question now is whether
such a simple ansatz has a natural realization.  It may be of interest to
note that in the discredited case of the 17-keV neutrino, the most probable
theoretical explanation was that a linear combination of $\nu_e$ and
$\nu_\tau$ pairs up with $\nu_\mu$ to form a pseudo-Dirac neutrino.\cite{16}

Since $m$ and $\Delta m_{12}^2$ should be of order 0.1 eV and
$\rm 10^{-10} eV^2$ respectively, the ratios $\epsilon_{1,2,3,4}/m$ are
of order $10^{-8}$.  Hence it is natural to assume as a first approximation
that $\epsilon_1 = \epsilon_2 = \epsilon_3 = \epsilon_4 = 0$.  This can be
achieved by the imposition of a discrete symmetry which is then softly
broken so that $\epsilon_{1,2,3,4}$ may acquire small nonzero values.
Since $m$ itself is already rather small, a natural explanation is that
of radiative generation.\cite{17}  In the following it will be shown how
everything can be done in the context of the recently proposed doublet
Majoron model.\cite{10,11}

If there is no $\nu_S$ and ${\cal M}_\nu$ refers to the known three
light neutrinos, then they have no impact on the question of dark matter
in the Universe because the sum of their masses would be much less than 1 eV.
After the results of the Cosmic Background
Explorer (COBE),\cite{18} it is popularly assumed that the Universe contains
70\% cold dark matter and 30\% hot dark matter.\cite{19}  The latter could
be neutrinos, but the sum of their masses has to be about 7 eV.  Implications
of this assumption on the neutrino mass matrix have been explored.\cite{20}
On the other hand, it is also possible that the Universe contains 100\%
cold dark matter and the COBE results are explained by a late decaying
particle,\cite{11,21,22,23,24} the prime candidate being $\nu_\tau$, but
its mass should be a few MeV.  There is actually another good reason for
a $\nu_\tau$ of this mass.  Its Yukawa coupling would then be large enough to
allow for the possible generation of the observed baryon-number asymmetry of
the Universe during the electroweak phase transition from the spontaneous
breaking of lepton-number conservation.\cite{25}  This mechanism requires
a detailed understanding of transmission through and reflection off bubble
walls, and is under active investigation.\cite{26}

The recently proposed doublet Majoron model\cite{10,11} provides a
natural framework for both electroweak baryogenesis and cold dark matter.
Since $m_{\nu_\tau}$ is a few MeV in this case, the mass matrix ${\cal M}_\nu$
of Eq. (1) should now be interpreted as representing $\nu_e, \nu_\mu$, and
$\nu_S$, the last being a singlet neutrino, each having lepton
number $L = 1$.  Note that in this model,\cite{10,11} lepton number
corresponds to
a conserved global U(1) symmetry above the energy scale of electroweak
symmetry breaking.  It is broken spontaneously together with the SU(2)
$\times$ U(1) gauge symmetry necessarily and a lepton asymmetry of the
Universe is created which gets converted into a baryon asymmetry through
sphalerons.\cite{25}  The massless Goldstone boson associated with the
spontaneous breaking of $L$ is called the Majoron.  The massive $\nu_\tau$'s
annihilate into Majorons very quickly in this model so that the $\nu_\tau$
contribution to the energy density of the Universe at the time of
nucleosynthesis is negligible.  On the other hand, $\nu_\tau$ decays
rather slowly and as the Universe expands, it eventually becomes dominant,
but only until it finally decays away into Majorons and other light neutrinos.
This scenario is thus very much suited for the radiative generation
of Majorana neutrino masses\cite{17} because lepton number is already
assumed to be spontaneously broken.

In addition to all the particles of the standard model, let there be one
light singlet neutrino $\nu_{SL}$ with $L = 1$, one heavy neutral singlet
fermion $N_R$ with $L = 0$, and two scalar doublets $\Phi_{1,2}
= (\phi_{1,2}^+, \phi_{1,2}^0)$ with $L = \mp 1$.  To obtain $\epsilon_1
= \epsilon_2 = \epsilon_3 = \epsilon_4 = 0$ in ${\cal M}_\nu$, assume
a discrete $Z_3$ symmetry such that $(\nu_e, e)_L, (\nu_\mu, \mu)_L, e_R,
\mu_R$ transform as $\omega$, whereas $(\nu_\tau, \tau)_L, \tau_R, \nu_{SL},
N_R$ transform as $\omega^2$, with $\omega^3 = 1$.  To obtain radiative
neutrino masses, assume the existence of three charged scalar singlets
$\eta^-_{0,1,2}$ with $L = 0,1,2$ respectively.  All scalar particles are
assumed to be trivial under $Z_3$.  As a result, the $\nu_e - \nu_S$ and
$\nu_\mu - \nu_S$ mass terms are generated in one loop as shown in Fig. 1,
but all other entries of ${\cal M}_\nu$ remain zero.  Specifically,
\begin{eqnarray}
m \cos \theta &=& {{f_{e\tau} m_\tau f_{\tau S}} \over {16 \pi^2}}
{{v_1^2 v_0^2} \over M^4}, \\ m \sin \theta &=& {{f_{\mu\tau} m_\tau
f_{\tau S}} \over {16 \pi^2}} {{v_1^2 v_0^2} \over M^4},
\end{eqnarray}
where $v_{0,1}$ are the vacuum expectation values of $\phi^0_{0,1}$,
and $M$ is an effective mass of the $\eta$'s in the loop.  However,
${\cal M}_\nu$ is only a submatrix of a larger $5 \times 5$ matrix
containing also $\nu_\tau$ and $N$.  Assuming a heavy Majorana mass
for $N$ (which breaks $Z_3$ softly), $\nu_\tau$ gets a seesaw mass due
to its coupling to $N$ via $\phi_1^0$.  The effective $4 \times 4$ mass
matrix spanning $\nu_e, \nu_\mu, \nu_S$, and $\nu_\tau$ is then given by
\begin{equation}
{\cal M}'_\nu = \left[ \begin{array} {c@{\quad}c@{\quad}c@{\quad}c} 0 &
0 & m \cos \theta & m' \cos \theta' \\ 0 & 0 & m \sin \theta & m' \sin
\theta' \\ m \cos \theta & m \sin \theta & 0 & 0 \\ m' \cos \theta' &
m' \sin \theta' & 0 & m_{\nu_\tau} \end{array} \right].
\end{equation}
The $\nu_e - \nu_\tau$ and $\nu_\mu - \nu_\tau$ mass terms are
also radiatively induced in one loop as in Fig. 1, but with $\nu_S$
replaced by $\nu_\tau$ and $\eta_0^-$ by $\phi_0^-$.  As a result,
\begin{eqnarray}
m' \cos \theta' &=& {{f_{e\tau} (m_\tau^2 - m_e^2)} \over {16 \pi^2}}
{{\Lambda v_1^2} \over M^4}, \\ m' \sin \theta' &=& {{f_{\mu\tau}
(m_\tau^2 - m_\mu^2)} \over {16 \pi^2}} {{\Lambda v_1^2} \over M^4},
\end{eqnarray}
where $\Lambda$ is the cubic $\phi_0^+ \phi_1^0 \eta_1^-$ coupling.
Comparing Eqs. (13) and (14) to Eqs. (10) and (11), it is clear that
$\theta \simeq \theta'$, and $\sin^2 2 \theta = 0.5$ is obtained if
$f_{\mu\tau}/f_{e\tau} = 0.4$.  Using $M = 1$ TeV, $v_0 = 245$ GeV, and
$v_1 = 22$ GeV, a value of 0.1 eV for $m$ is also obtained if
$\sqrt {f_{e\tau}^2 + f_{\mu\tau}^2} = 0.01$ and $f_{\tau S} = 0.03$.
Because of mixing with $\nu_\tau$, the effective ${\cal M}_\nu$ of Eq. (1)
now has
\begin{eqnarray}
&~& \epsilon_1 \simeq m'^2 \cos^2 \theta / m_{\nu_\tau}, ~~~~ \epsilon_2
\simeq m'^2 \sin^2 \theta / m_{\nu_\tau}, \\ &~& \epsilon_3 = 0, ~~~~
\epsilon_4 \simeq m'^2 \sin \theta \cos \theta / m_{\nu_\tau}.
\end{eqnarray}
Therefore,
\begin{equation}
\Delta m_{12}^2 \simeq 2 m m'^2 / m_{\nu_\tau}.
\end{equation}
Using $\Lambda = 400$ GeV, a value of about 0.04 eV for $m'$ is obtained.
Hence $\Delta m_{12}^2 \simeq 10^{-10}{\rm eV}^2$ if $m_{\nu_\tau} \simeq
3$ MeV.  These numbers clearly demonstrate that a natural radiative
realization of ${\cal M}_\nu$ is possible for a successful explanation
of the solar and atmospheric neutrino deficits.  It should be mentioned
that ${\cal M}'_\nu$ of Eq. (12) has also been obtained with a Dirac
seesaw mechanism in a recently proposed singlet-triplet Majoron
model.\cite{24}

Consider now the decay of $\nu_\tau$ in the present model.  It proceeds
via the mixing of $\nu_e$ and $\nu_\mu$ with $\nu_\tau$ in
${\cal M}'_\nu$ which is $m'/m_{\nu_\tau}$.  The rate is given by\cite{11}
\begin{equation}
\Gamma = {{m'^2 m_{\nu_\tau}} \over {64 \pi v_1^2}}.
\end{equation}
For $m_{\nu_\tau} = 3$ MeV, the $\nu_\tau$ lifetime is then
about $1.3 \times 10^4$ seconds, which is within the required
range for a successful explanation of the COBE data in the case of 100\%
cold dark matter.\cite{11}  This is a remarkable correlation between the
constraint of cosmology and that of solar and atmospheric neutrino
data.

The singlet neutrino $\nu_S$ is not inert, but because of the discrete
$Z_3$ symmetry, its only interaction at tree level with the other leptons
is given by $f_{\tau S} \overline \tau_R \nu_S \eta_0^- + H.c.$  Hence its
effect on all known leptonic processes is easily shown to be negligible
for $f_{\tau S} = 0.03$ and $m_\eta = 1$ TeV.  It decouples from other
light particles in the early Universe when the $\tau$ does.  Hence its
contribution to the energy density at the time of nucleosynthesis is also
negligible.  Since $m_{12}$ is of order $10^{-10} {\rm eV}^2$, the
oscillation time between $\nu_e$ and $\nu_S$ is about $10^2$ seconds.
This is long enough also for $\nu_S$ not to be a factor in nucleosynthesis.
In fact, the contributing light degrees of freedom in this model, not
counting the photon, consists of only $\nu_e, \nu_\mu$, and the Majoron.
Hence the effective number of neutrinos $N_\nu$ is only 2.6, below the
standard upper bound of 3.3\cite{27} or the more recently proposed
3.04\cite{28}.

Since $\eta_2$ couples to the leptons via the interactions $(\nu_e \tau_L -
e_L \nu_\tau) \eta_2^+$ and $(\nu_\mu \tau_L - \mu_L \nu_\tau) \eta_2^+$,
there are additional contributions to leptonic processes.  For example,
$\mu \rightarrow e \overline \nu_e \nu_\mu$ decay is accompanied by
$\mu \rightarrow e \overline \nu_\tau \nu_\tau$ but the latter is only
of order $10^{-6}G_F$ in strength.  Similarly, $\mu \rightarrow e \gamma$
and $\nu_\tau \rightarrow \overline \nu_e \gamma + \overline \nu_\mu \gamma$
have branching fractions of order $10^{-14}$, and $\nu_\tau \rightarrow
e^- e^+ \overline \nu_e$ is even more negligible.  Hence the standard
low-energy weak-interaction phenomenology is not affected.  A second
comment involves $CP$ nonconservation.  In the above, since only one $N_R$
is assumed, the $\nu_\tau$ Yukawa coupling to $\phi_1^0$ can be
chosen real.  Nevertheless, $CP$ nonconserving couplings do exist in the
Higgs sector which may or may not be sufficient for electroweak baryogenesis.
If not, an easy remedy is to add one more $N_R$, then a $CP$ nonconserving
phase will show up explicitly in the $\nu_\tau$ Yukawa coupling.

In conclusion, it has been shown in this paper that a simple ansatz for
the neutrino mass matrix, {\it i.e.} ${\cal M}_\nu$ of Eq. (1), works
very well as an explanation of the present observed solar and atmospheric
neutrino deficits.  It is also naturally realized by a radiative mechanism
based on the spontaneous breaking of lepton number.  This has the advantage
of incorporating electroweak baryogenesis and allowing the missing mass of
the Universe to be all cold dark matter.  The key is for $\nu_\tau$ to be
a few MeV in mass and to decay late enough to delay the ultimate time of
matter-radiation equality in the early Universe.  This has been accomplished
in a previously proposed doublet Majoron model,\cite{10,11} which is now
extended to include a singlet neutrino $\nu_{SL}$ with $L = 1$ and three
charged scalar singlets together with a softly broken discrete $Z_3$
symmetry, resulting in an effective ${\cal M}_\nu$ exactly of the right
form.  Because of the necessity of maximum mixing, only the vacuum
oscillation solution of the solar neutrino deficit is applicable in
this scenario.  However, the numbers turn out to be just right for the
$\nu_\tau$ lifetime.  Specifically, $m \simeq 0.1$ eV from the
atmospheric data, $m m'^2 / m_{\nu_\tau} \simeq 10^{-10} {\rm eV}^2$
from the solar data, and $m_{\nu_\tau} \sim $ few MeV, $m' / m_{\nu_\tau}
\sim 10^{-8}$ from cosmology.

{\it Note Added.} If there are no neutrinos beyond $\nu_e$, $\nu_\mu$, and
$\nu_\tau$, it is still possible to obtain ${\cal M}_\nu$ of Eq. (1)
radiatively.  Since a Majoron is not required, lepton number will now
be assumed to be broken by explicit soft terms.  In particular,
the cubic term $\eta_2^- \phi_0^+ \phi_1^0$ is allowed.  Hence the
$\nu_e - \nu_\tau$ and $\nu_\mu - \nu_\tau$ entries are radiatively
generated in one loop, but the other entries remain zero.  Now let there
be a doubly charged singlet scalar $\sigma^{--}$ with lepton number $L = 2$
and which transforms as $\omega$ under $Z_3$, then the interaction
$\sigma^{++} \tau_R \tau_R$ is allowed (but not with $\tau_R$ replaced by
$e_R$ or $\mu_R$).  Let there also be the cubic term $\sigma^{++} \eta_2^-
\eta_2^-$ which breaks both $L$ and $Z_3$, then these other entries also
become nonzero in two loops.\cite{17}  Hence the desired form of
the $3 \times 3$ ${\cal M}_\nu$ is again realized radiatively.

\newpage
\begin{center} {ACKNOWLEDGEMENT}
\end{center}

This work was supported in part by the U. S. Department of Energy under
Grant No. DE-FG03-94ER40837.

\bibliographystyle{unsrt}

\begin{thebibliography}{99}
\bibitem{1} R. Davis, Jr. {\it et al.}, Ann. Rev. Nucl. and Part. Sci.
{\bf 39}, 467 (1989).
\bibitem{2} K. S. Hirata {\it et al.}, Phys. Rev. Lett. {\bf 63}, 16
(1989); {\bf 65}, 1297 (1990); {\bf 66}, 9 (1991).
\bibitem{3} A. I. Abazov {\it et al.}, Phys. Rev. Lett. {\bf 67}, 3332
(1991).
\bibitem{4} P. Anselmann {\it et al.}, Phys. Lett. {\bf B314}, 445 (1993);
{\bf B327}, 377 (1994).
\bibitem{5} R. Becker-Szendy {\it et al.}, Phys. Rev. D {\bf 46}, 3720 (1992).
\bibitem{6} K. S. Hirata {\it et al.}, Phys. Lett. {\bf B280}, 146 (1992);
Y. Fukuda {\it et al.}, {\it ibid.} {\bf B335}, 237 (1994).
\bibitem{7} For solar neutrinos, see for example D. Harley, T. K. Kuo,
and J. Pantaleone, Phys. Rev. {\bf D47}, 4059 (1993).
\bibitem{8} For atmospheric neutrinos, see for example J. Pantaleone,
Phys. Rev. {\bf D49}, R2152 (1994).
\bibitem{9} M. Gell-Mann, P. Ramond, and R. Slansky, in {\it Supergravity},
edited by P. van Nieuwenhuizen and D. Z. Freedman (North-Holland,
Amsterdam, 1979), p. 315; T. Yanagida, in {\it Proceedings of the Workshop
on the Unified Theory and the Baryon Number in the Universe}, Tsukuba,
Ibaraki, Japan, 1979, edited by O. Sawada and A. Sugamoto (KEK, Tsukuba,
Japan, 1979).
\bibitem{10} H. Kikuchi and E. Ma, Phys. Lett. {\bf B335}, 444 (1994).
\bibitem{11} H. Kikuchi and E. Ma, Phys. Rev. {\bf D} (Rapid Communication),
to be published.
\bibitem{12} See for example W. Frati {\it et al.}, Phys. Rev. {\bf D48},
1140 (1993).
\bibitem{13} See for example V. Barger, R. J. N. Phillips, and K. Whisnant,
Phys. Rev. Lett. {\bf 69}, 3135 (1992).
\bibitem{14} See for example A. Acker, A. B. Balantekin, and F. Loreti,
Phys. Rev. {\bf D49}, 328 (1994).
\bibitem{15} S. P. Mikheyev and A. Yu. Smirnov, Yad. Fiz. {\bf 42}, 1441
(1985) [Sov. J. Nucl. Phys. {\bf 42}, 913 (1985)]; Nuovo Cimento {\bf 9C},
17 (1986); L. Wolfenstein, Phys. Rev. {\bf D17}, 2369 (1978).
\bibitem{16} See for example E. Ma, Phys. Rev. Lett. {\bf 68}, 1981 (1992).
\bibitem{17} For a brief review, see for example K. S. Babu and E. Ma,
Mod. Phys. Lett. {\bf A4}, 1975 (1989).


\bibitem{18} G. F. Smoot {\it et al.}, Astrophys. J. {\bf 396}, L1 (1992).
\bibitem{19} Q. Shafi and F. W. Stecker, Phys. Rev. Lett. {\bf 53}, 1292
(1984); E. L. Wright {\it et al.}, Astrophys. J. {\bf 396}, L13 (1992).
\bibitem{20} D. O. Caldwell and R. N. Mohapatra, Phys. Rev. {\bf D48},
3259 (1993).
\bibitem{21} S. Dodelson, G. Gyuk, and M. Turner, Phys. Rev. Lett. {\bf 72},
3754 (1994).
\bibitem{22} H. B. Kim and J. E. Kim, Seoul National University Report No.
SNUTP-94-48 (1994); E. J. Chun, ICTP Trieste Report No. IC/94/306 (1994).
\bibitem{23} R. N. Mohapatra and S. Nussinov, University of Maryland Report
No. UMD-PP-95-21 (1994).
\bibitem{24} A. S. Joshipura and J. W. F. Valle, University of Valencia
Report No. FTUV/94-46 (1994).
\bibitem{25} A. Cohen, D. Kaplan, and A. Nelson, Phys. Lett. {\bf B245},
561 (1990); Nucl. Phys. {\bf B349}, 727 (1991).
\bibitem{26} See for example M. Joyce, T. Prokopec, and N. Turok, Princeton
University Reports PUPT-1437, PUPT-1495, PUPT-1496, PUPT-1497.
\bibitem{27} T. Walker {\it et al.}, Astrophys. J. {\bf 376}, 51 (1991).
\bibitem{28} P. Kernan and L. Krauss, Phys. Rev. Lett. {\bf 72}, 3309 (1994).
\end{thebibliography}

\vspace{0.3in}
\begin{center}{FIGURE CAPTION}
\end{center}

Fig. 1. One-loop radiative $\nu_e - \nu_s$ mass due to the spontaneous
breaking of lepton number.

\end{document}